\documentclass[aps,prl,preprint,groupedaddress,12pt]{revtex4-1}

\usepackage[T1]{fontenc}
\usepackage[latin9]{inputenc}
\usepackage[a4paper]{geometry}
\usepackage{amsmath}
\usepackage{graphicx}
\usepackage{amssymb}
\usepackage{esint}
\usepackage{color}

\begin{document}

\title{Thermalization kinetics of light: From laser dynamics to equilibrium condensation of photons}

\author{Julian Schmitt}
\author{Tobias Damm}
\author{David Dung}
\author{Frank Vewinger}
\author{Jan Klaers}
\altaffiliation{Present address: Institute for Quantum Electronics, ETH Z\"urich, Auguste-Piccard-Hof 1, 8093 Z\"urich, Switzerland}
\author{Martin Weitz}
\affiliation{Institut f\"ur Angewandte Physik, Universit\"at Bonn, Wegelerstr.
8, 53115 Bonn, Germany}

\pacs{03.75.Kk,42.50.Gy,67.85.Hj}

\begin{abstract}
We report a time-resolved study of the thermalization dynamics and the lasing to photon Bose-Einstein condensation crossover by in-\textit{situ} monitoring the photon kinetics in a dye microcavity. When the equilibration of the light to the dye temperature by absorption and re-emission is faster than photon loss in the cavity, the optical spectrum becomes Bose-Einstein distributed and photons accumulate at low-energy states, forming a Bose-Einstein condensate. The thermalization of the photon gas and its evolution from nonequilibrium initial distributions to condensation is monitored in real-time. In contrast, if photons leave the cavity before they thermalize, the system operates as a laser.
\end{abstract}
\maketitle
Thermal equilibrium, the state of minimum free energy for a system coupled to a thermal bath, is a cornerstone of statistical mechanics, with its predictions including ordering phenomena at a phase transition~\cite{huang_statistical_1987}. One intriguing example is Bose-Einstein condensation (BEC), where a gas of bosons macroscopically occupies the ground state below a critical temperature~\cite{bose_plancks_1924,einstein_quantentheorie_1925}. This phenomenon has been studied extensively with cold atomic gases~\cite{cornell_nobel_2002,ketterle_nobel_2002,pethick_bose-einstein_2002}, polaritons both freely flowing and trapped~\cite{deng_exciton-polariton_2010,balili_bose-einstein_2007,cristofolini_optical_2013}, and more recently also with photons confined in a dye-filled microscopic optical cavity~\cite{klaers_bose-einstein_2010,marelic_experimental_2015}. In contrast to blackbody radiation, where the photon number is not conserved upon variation of the temperature, photon energies here are restricted to values well above thermal energy by a low-frequency cutoff, introducing a non-vanishing chemical potential for the photons. In earlier work, we have observed a thermalized two-dimensional photon gas~\cite{klaers_thermalization_2010}, the emergence of Bose-Einstein condensation~\cite{klaers_bose-einstein_2010} and the condensate second-order coherence~\cite{schmitt_observation_2014}.

In general however, driving and loss can force a system away from thermal equilibrium~\cite{haken_cooperative_1975,muller_engineered_2012,kasprzak_formation_2008,wouters_spatial_2008,kirton_nonequilibrium_2013}. The perhaps most prominent example for the emergence of an ordered state in nonequilibrium physics is the laser, with macroscopically occupied modes of arbitrary energy~\cite{siegman_lasers_1986}. Here, both the state of the light field and that of the amplifying medium are far from equilibrium, allowing for inversion and gain. Whether a system that is pumped and exhibits losses as well as thermalization should rather be described in terms of the BEC equilibrium framework or laser theory has been discussed for exciton-polaritons~\cite{kasprzak_formation_2008,wouters_spatial_2008,deng_quantum_2006}, and also evolution from photon to polariton lasing has been studied~\cite{kammann_crossover_2012}. Kirton and Keeling provided a nonequilibrium theoretical model of the transition from photon condensation to laser-like states~\cite{kirton_nonequilibrium_2013,kirton_thermalization_2015}, and other work has proposed tests of photon equilibration processes based on the fluctuation-dissipation theorem~\cite{chiocchetta_laser_2015}. Further, the condensation dynamics of classical optical waves was studied~\cite{sun_observation_2012}. Due to the ability to tune both photon loss as well as the equilibration time, photons in an optical dye microcavity present an excellent system to study the crossover from lasing to Bose-Einstein condensation, phenomena both showing spontaneous symmetry breaking, see e.g. Refs.~\cite{haken_cooperative_1975,scully} for early discussions of analogies.

As the here studied system operates in the weak coupling regime~\cite{angelis_microcavity_2000}, consider a rate equation model for photons, where dye molecules are modeled as rovibrationally broadened two-level systems, and the cavity is described by a loss rate $\Gamma(\omega)$. The net temporal variation of the photon number at frequency $\omega$ is
\begin{eqnarray}
\frac{\textrm dn(\omega)}{\textrm dt}&=&M_\uparrow A(\omega)+M_\uparrow \hat{B}_{21}(\omega)n(\omega)\nonumber\\
&&-M_\downarrow \hat{B}_{12}(\omega) n(\omega)-\Gamma(\omega)n(\omega),
\label{rateequation}
\end{eqnarray}
where $\hat{B}_{12}(\omega)$ and $\hat{B}_{21}(\omega)$ denote Einstein rate coefficients for absorption and stimulated emission respectively, and $A(\omega)$ accounts for spontaneous emission. The Einstein $A$-$B$ relation here takes the form $A(\omega)/\hat{B}_{21}(\omega)=g(\omega)$, with $g(\omega)$ as the mode degeneracy~\cite{supplemental}. Further, $M_\downarrow$ and $M_\uparrow$ are the populations in ground and electronically excited molecular states, respectively, whose ratio is determined by external pumping. Usual laser equations follow from Eq.~(\ref{rateequation}), when assuming an open system where spontaneous emission is not recaptured, and lasing sets in when the gain at a particular frequency overcomes the loss~\cite{siegman_lasers_1986}. On the other hand, in the limit of an ideal "photon box" ~\cite{bohr_discussion_1988} with $\Gamma(\omega)=0$, the rate equation model for the closed system yields a steady-state photon distribution approaching a Bose-Einstein distribution~\cite{klaers_statistical_2012,klaers_boseeinstein_2011}, in analogy to e.g. the kinetics of atomic Bose gases~\cite{lee_quantum_2000}. Because thermalization of the photon gas does not occur by photon-photon interactions but to an external reservoir, it is required that the dye reservoir fulfills the Boltzmann-type scaling predicted by the Kennard-Stepanov relation $\hat{B}_{21}(\omega)/\hat{B}_{12}(\omega)\propto \exp(-\hbar\omega/k_\textrm B T)$, related to both the ground and electronically excited rovibrational manifold each being in equilibrium due to frequent (fs timescale) collisions of solvent molecules with the dye~\cite{lakowicz_principles_1999}. Photons thermalize to the rovibrational temperature of the dye, room temperature ($T=300$~K) in our case. 

\begin{figure}[t]
\includegraphics[width=0.5\columnwidth]{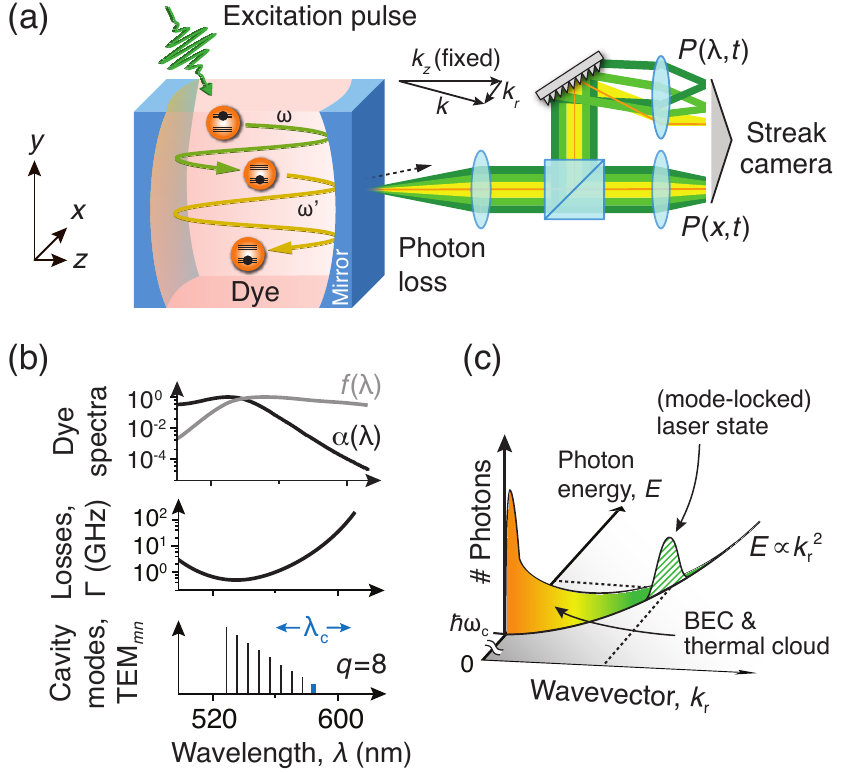}
\caption{(a) Optical microresonator filled with dye solution. While trapped in the resonator, photons thermalize by dye absorption re-emission cycles. Light leaking from the cavity is monitored spatially or spectrally using a streak camera. (b) Measured, normalized spectral profiles of dye absorption $\alpha(\lambda)$, fluorescence $f(\lambda)$ (top panel), and cavity loss rate $\Gamma(\lambda)$ (middle panel). The bottom panel schematically shows the cavity transverse mode structure with the cutoff wavelength $\lambda_\textrm c$, indicating the linearly increasing degeneracy of the excited transverse modes at lower wavelengths. (c) Quadratic photon dispersion in the cavity along with a Bose-Einstein condensed (filled) and a nonequilibrium spectral photon distribution (hatched).
\label{fig1}}
\vspace{-4mm}
\end{figure}

The experimental scheme used to study photon gas dynamics is shown in Fig.~\ref{fig1}(a). Photons are confined in a microcavity~\cite{klaers_bose-einstein_2010,schmitt_observation_2014}, which consists of two spherically curved, highly reflecting mirrors spaced $1.6~\mu\textrm{m}$ apart, and is filled with rhodamine dye solution. The small mirror separation imposes a frequency spacing of longitudinal resonator modes that exceeds thermal energy ($\simeq 1/40~\textrm{eV}$ at room temperature, corresponding to $6\times 10^{12}~\textrm{Hz}$ in frequency units) and is comparable to the spectral width of the dye emission. Thus, only photons of a fixed longitudinal mode number ($q=8$) populate the cavity. This effectively restricts the energies to be above a minimum cutoff of $\hbar\omega_\textrm c = 2.1~\textrm{eV}$, the eigenenergy of the corresponding TEM$_{00}$ mode. Higher order transverse modes lie at higher energies, with a quadratic relation between energy and transverse momentum, see Fig.~\ref{fig1}(c). For a sufficiently long storage time in the optical resonator, the two transverse modal quantum numbers thermalize by repeated absorption re-emission processes to the temperature $T$ of the dye solution. The photon energies will then be distributed by an amount ${\sim}k_\textrm{B}T$ above the cavity cutoff. One can show that the system is formally equivalent to a two-dimensional, harmonically trapped gas of massive bosons~\cite{klaers_thermalization_2010}, and above a critical photon number Bose-Einstein condensation to the ground mode is possible~\cite{klaers_bose-einstein_2010}. Photons are injected into the microresonator by pumping the dye with a laser beam near $532~\textrm{nm}$ and of $15~\textrm{ps}$ pulse length. By changing pump focal diameter and transverse position, different spatial distributions of initial molecular electronic excitations are realized. The time evolution of the photon distribution following the dye excitation is analyzed spatially and spectrally by imaging light transmitted through one of the cavity mirrors either directly or spectrally dispersed onto a streak camera (Fig.~\ref{fig1}(a)). Both photon loss and thermalization rates can be tuned by variation of the cavity cutoff, as the dye absorption as well as the cavity loss rate due to mirror transmission are wavelength dependent, see Fig.~\ref{fig1}(b).

\begin{figure}[h]
\includegraphics[width=0.5\columnwidth]{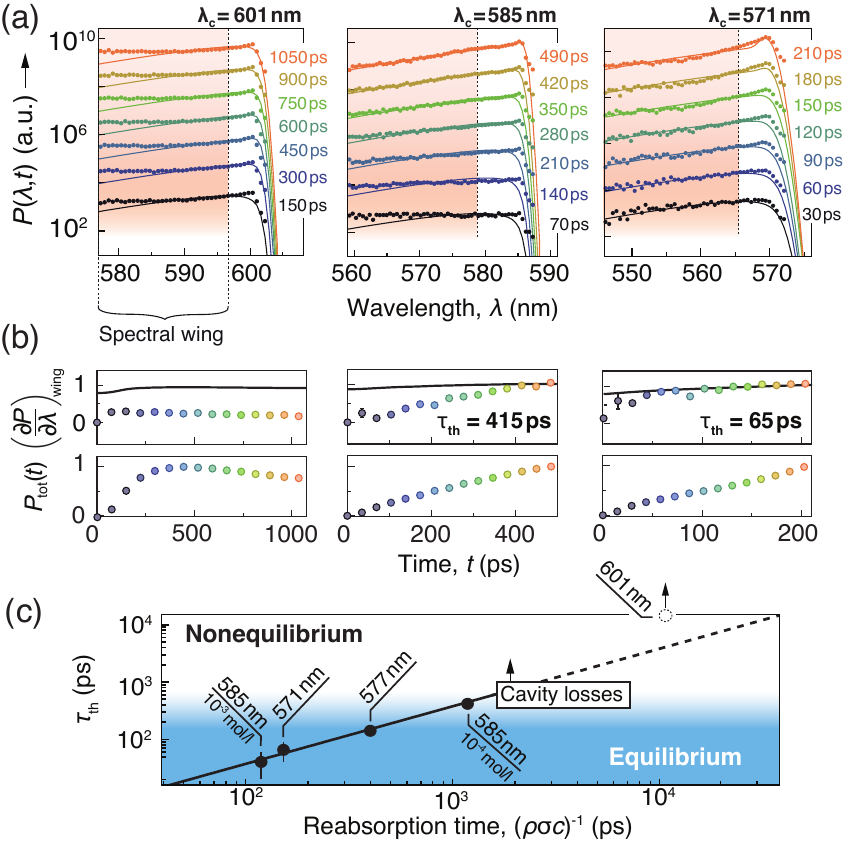}
\caption{(a) Spectral distribution for different times after the pump pulse (dots, vertically shifted) for different cutoff wavelengths $\lambda_\textrm c$, along with 300~K Bose-Einstein distributions (solid lines). The low-wavelength part of the experimental spectra (shaded) is fitted with a linear function, and in (b) (top panel) the evolution of the slope of the thermal wings (circles) is compared to that of a fully thermalized spectrum with corresponding chemical potential (solid line). For $601~\textrm{nm}$ (left) no thermalization is observed, while for a cutoff closer to the dye absorption maximum photons thermalize within the cavity lifetime (middle and right). The total cavity emission (bottom panel) yields the timescales for chemical equilibration between photons and dye excitations $\tau_\textrm{ch}=\left\{450~\textrm{ps};560~\textrm{ps};>350~\textrm{ps}\right\}$ (note the different scales on the time axis). (c) Observed photon gas thermalization time $\tau_\textrm{th}$ (data points) versus dye reabsorption time compared to theory (solid line), showing the transition from nonequilibrium conditions to thermalization for larger dye absorption and longer cavity lifetimes~\cite{supplemental}. Experimental parameters: $\lambda_\textrm c=\left\{601~\textrm{nm};585~\textrm{nm};577~\textrm{nm};571~\textrm{nm}\right\}$ (corresponding cavity photon lifetimes $\left\{26~\textrm{ps};162~\textrm{ps};331~\textrm{ps};540~\textrm{ps}\right\}$) at $\rho=10^{-4}~\textrm{mol/l}$, and $\lambda_\textrm c=585~\textrm{nm}$ at $\rho=10^{-3}~\textrm{mol/l}$.
\vspace{-4mm}
\label{fig2}}
\end{figure}

First we have studied the thermalization dynamics of the photon gas near the BEC threshold ($N_\textrm{c}\approx 90,000$), for which the dye was excited spatially homogeneous with a large ($\approx0.5$~mm diameter) pump beam. Figure \ref{fig2}(a) gives measured spectra for advancing times after the pump pulse, for comparison together with the corresponding 300~K Bose-Einstein distributions~\cite{supplemental}. The shown spectral range of $\Delta\lambda=25~\textrm{nm}$, corresponding to $\Delta E\simeq 3.5k_\textrm B T$, for a thermalized spectrum contains at least 93\% of all photons inside the microcavity. The data shown on the left hand side was recorded with a cutoff wavelength $601~\textrm{nm}$, for which the high-energy (low-wavelength) tail throughout the measurement remains far from a thermalized distribution. In this wavelength range small dye absorption and high cavity loss (see Fig.~\ref{fig1}(b)) prevent equilibration of photons within the cavity lifetime. The data recorded with cutoff wavelengths of $585~\textrm{nm}$ and $571~\textrm{nm}$, respectively (middle and right), show proceeding thermalization, as expected from the here larger dye absorption and lower cavity loss. To quantify this, the circles in Fig.~\ref{fig2}(b) show the high-energy slope of each spectrum. For the data sets with larger dye reabsorption (middle and right), the expected slope for a fully thermalized ensemble is approached, and one can determine the corresponding thermalization times $\tau_\textrm{th}\simeq 415~\textrm{ps}$ and $65~\textrm{ps}$, respectively. In the latter case, this is one order of magnitude below the corresponding cavity lifetime of $540~\textrm{ps}$ so that thermalization of the distribution is expected. The quoted values for the thermalization time give the time at which a residual deviation of 1\% from the slope of an equilibrium spectrum is reached. The bottom panels of Fig.~\ref{fig2}(b) give the temporal variation of the total cavity emission, which increases until chemical equilibrium between the number of photons and dye electronic excitations is reached. Figure~\ref{fig2}(c) shows the observed thermalization times $\tau_\textrm{th}$ versus dye reabsorption time $(\rho\sigma(\lambda)c)^{-1}$, where $\rho$ denotes dye concentration, $\sigma(\lambda)$ the dye absorption cross section, and $c$ the speed of light (for the additional data see~\cite{supplemental}). We observe a linear scaling (solid line) following the relation $\tau_\textrm{th}=0.37(5)\cdot (\rho\sigma(\lambda)c)^{-1}$. This demonstrates improved thermal contact for increasing dye absorption, and photons thermalize roughly as fast as they are reabsorbed, except for $\tau_\textrm{th} > 500~\textrm{ps}$, where the finite cavity lifetime prevents thermalization of the photon gas.

\begin{figure}[b]
\includegraphics[width=0.5\columnwidth]{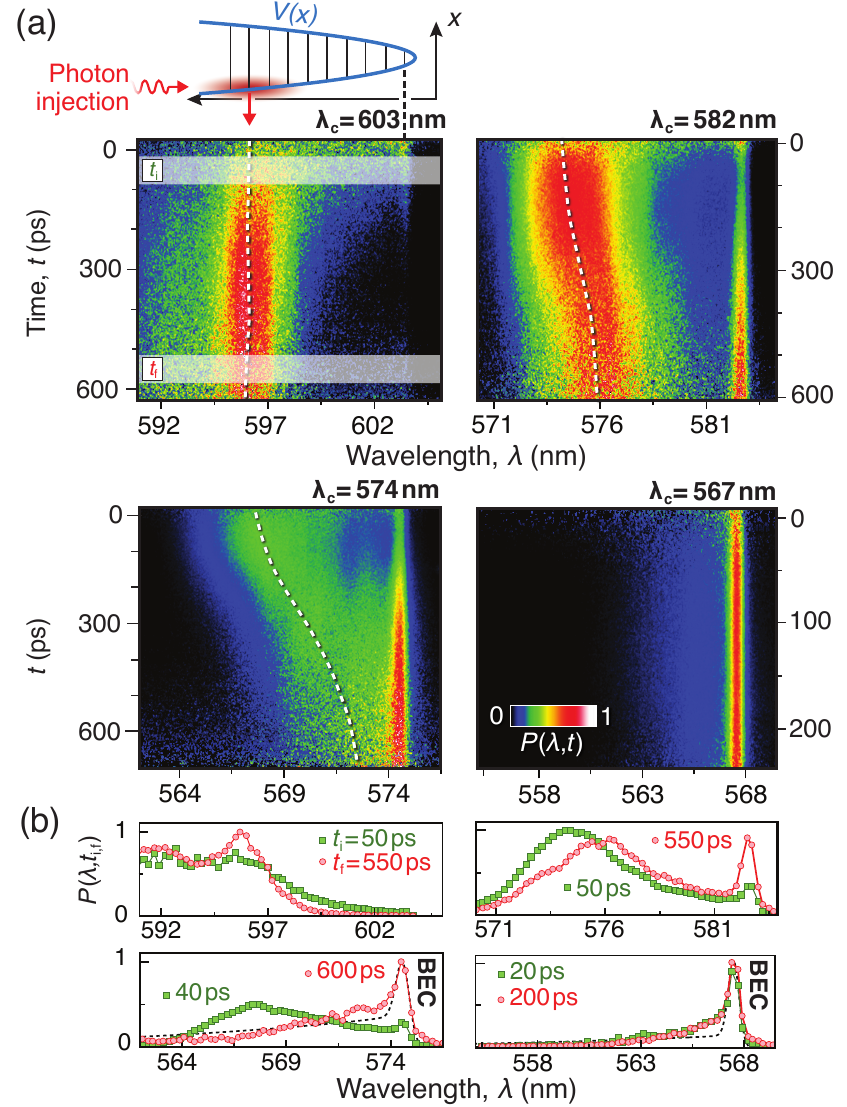}
\caption{(a) Spectral evolution of the photon gas dynamics (line-normalized) for off-center pumping with a $80~\mu\textrm{m}$ diameter pump spot $160~\mu\textrm{m}$ offset from the cavity axis for different cutoff settings. The off-center pumping locally inverts the dye molecules, which decay by stimulated emission of photons into highly excited transverse modes overlapping with the pump region. The blue line on top indicates the harmonic trap potential. For a cutoff wavelength far red from the dye absorption peak no thermalization is observed (top, left). Photons partially thermalize to lower energies for larger dye absorption (top, right) and finally collapse into a bimodal distribution with a Bose-Einstein condensate peak (bottom graphs, with $\lambda_\textrm c=574~\textrm{nm}$ and $567~\textrm{nm}$, corresponding to $(\rho\sigma(\lambda_\textrm c)c)^{-1}=96~\textrm{ps}$ and $33~\textrm{ps}$ respectively). As an indication for the redistribution towards longer wavelength, the dashed white line illustrates the evolution of the maximum of the emission not in the ground mode. (b) Spectra measured shortly after photon injection (green squares) and near the end of the streak camera time trace (red circles), averaged over the shaded areas in (a). $\rho=2.5\cdot10^{-4}~\textrm{mol/l}$.
\label{fig3}}
\end{figure}

The above result can be explained using a simple estimate: Following the pump pulse, a spectrally nearly flat emission into the cavity occurs in the relevant wavelength range (see Fig.~\ref{fig1}(b), upper panel, graph for $f(\lambda)$ above ${\sim}540~\textrm{nm}$). From this distribution, the high-energy tail is more likely reabsorbed due to the near exponential scaling of the absorption cross section $\sigma(\lambda)$ (note that the Kennard-Stepanov relation fixes the ratio $f(\lambda)/\sigma(\lambda)$), leading to a mostly thermalized photon distribution in the cavity. From Eq.~(\ref{rateequation}) one can accordingly estimate a timescale for this process from the reabsorption rate of $\tau_\textrm{th}\simeq (\hat B_{12}(\lambda)M_\downarrow)^{-1}$, see also the Supplemental Material~\cite{supplemental}. For weak excitation $M_\downarrow\simeq M$, $\hat B_{12}(\lambda)=\sigma(\lambda)c/V_\textrm{eff}$ with the mode volume $V_\textrm{eff}$ and $\rho=M/V_\textrm{eff}$ this gives $\tau_\textrm{th}\simeq (\rho\sigma(\lambda)c)^{-1}$, which reproduces the experimentally observed scaling for the thermalization of the spectral distribution. The photon number at this point however is in general still far from its equilibrium value. Only after the emission and absorption rates have balanced, both the correct spectral distribution and photon number are established. The timescale for chemical equilibration strongly depends on the electronic excitation level, which tunes the stimulated emission rate~\cite{supplemental}. The existence of two timescales may also be understood from the grand-canonical nature of the reservoir, which serves both as a heat bath and a particle reservoir for the photon gas \cite{klaers_statistical_2012,schmitt_observation_2014}.

In further measurements, we prepare an initial state of cavity photons energetically removed from the ground mode and investigate the thermalization to the heat bath (Fig.~\ref{fig3}), taking advantage of the harmonic trapping potential imposed by the mirrors' curvature. Therefore, we use a pump beam focus spatially displaced from the trap center, leading to a far nonequilibrium distribution of molecular excitations and photon emission occurring into (transversally) excited modes overlapping with the pump focus. Figure~\ref{fig3}(a) shows the temporal evolution of the spectrum and Fig.~\ref{fig3}(b) corresponding spectra both soon (green squares) and at a later time (red circles) after the pump pulse. As the dye absorption is increased with shorter cutoff wavelengths, photons thermalize, and for $\lambda_\textrm{c}=574~\textrm{nm}$ and $567~\textrm{nm}$ (the two data sets shown in the bottom panels of Fig.~\ref{fig3}(a) and (b), respectively) a bimodal distribution with a BEC peak and a thermalized wing emerges. In the latter case extremely fast thermalization within roughly $10~\textrm{ps}$ results in Bose-Einstein condensed spectra almost immediately after excitation.

\begin{figure}[h]
\includegraphics[width=0.5\columnwidth]{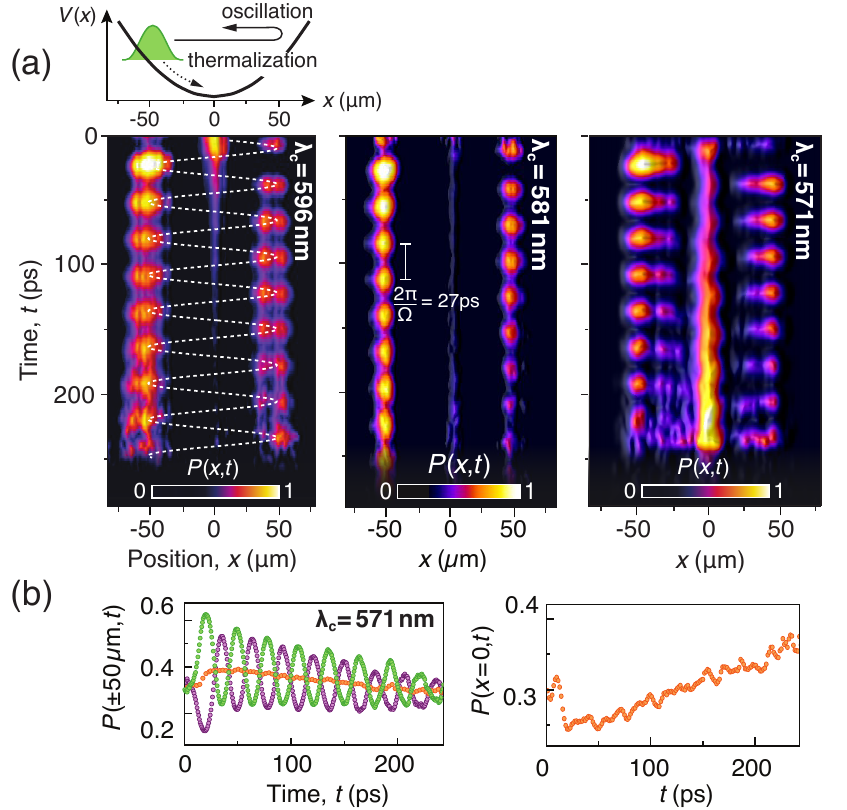}
\caption{(a) Mode-locked laser operation and photon condensate. Evolution of the spatial profile (line-normalized) of the radiation transmitted through one cavity mirror for three cutoff wavelengths. A pump beam with $27~\mu\textrm{m}$ diameter spatially displaced by $50~\mu\textrm{m}$ from the trap center excites an optical wave packet oscillating in the harmonic trapping potential (indicated on top). For the data recorded with largest cutoff wavelength (left) and correspondingly low dye absorption the oscillating mode-locked laser operation persists (classical oscillation indicated by dashed line), with the majority of the emission leaking out of the resonator at the turning points, while the rest is near the instrumental detection limit. In contrast, for larger reabsorption photons thermalize and a condensate in the trap center builds up (right). The middle graph depicts results for the crossover region between lasing and condensation. (b) (Left panel) Variation of the observed intensity in (a), (right), close to the oscillator reversal points near $x=\pm50~\mu\textrm{m}$ (purple and green respectively), their mean intensity (red) and (right panel) relative intensity in the condensate mode. $\rho=10^{-4}~\textrm{mol/l}$.
\label{fig4}}
\vspace{-4mm}
\end{figure}

To study the spatial dynamics of the photon gas after locally inverting the dye medium, we use a high intensity pump pulse focussed offset from the trap center. Figure~\ref{fig4}(a) gives false-color images obtained by directly imaging the cavity emission onto the streak camera, showing the intensity distribution along the $x$-axis versus time. Following the pump pulse, an optical wave packet is formed and oscillates back and forth in the trapping potential $V(x,y)$ imposed by the curved mirrors. The observed $27~\textrm{ps}$ oscillation period is in excellent agreement with the inverse trapping frequency of $2\pi/\Omega\simeq (37~\textrm{GHz})^{-1}$. Given the $4~\textrm{ns}$ upper electronic state spontaneous lifetime, the formation of the wave packet has to be induced by (fast) stimulated emission. Resonator modes with turning points near the position $(x,y)$ of the pump spot, i.e. near the tunable frequency $\omega=\omega_\textrm c + V(x,y)/\hbar$, experience maximum gain, as their spatial overlap with the region of inversion is largest. The oscillating wave packet is described by a coherent superposition of energy eigenstates and our system can be regarded as a mode-locked laser with an extremely high repetition rate. For the data recorded with longest cutoff wavelength and correspondingly small absorption and large losses shown on the left hand side of Fig.~\ref{fig4}(a), the emission persists in a mode-locked state. In contrast, for the data with shorter cutoff wavelength (middle and right), the oscillating wave packet gradually redistributes as photons thermalize, forming a BEC in the ground mode, in good qualitative agreement with a simplified theoretical model~\cite{supplemental}, and the spectral measurements of Fig.~\ref{fig3}. The measurement with largest contact to the dye (right) shows that thermalization from a laser-like state to a BEC takes place. The corresponding damping of the oscillation and the increasing condensate population is visible from Fig.~\ref{fig4}(b). 

To conclude, we have in-\textit{situ} monitored the photon gas kinetics in a dye microcavity and in particular observed a crossover between driven-dissipative system laser dynamics and a thermalized gas, with the steady-state being determined by energetics. While losses stabilize a nonequilibrium lasing operational mode, when photons remain long enough in the resonator to thermalize they accumulate at low energy states and form a Bose-Einstein condensate. In the latter case, the system is in thermal equilibrium in the same sense as a cloud of ultracold atoms is~\cite{pethick_bose-einstein_2002}.

Perspectives of this work include the investigation of thermalized photon gases in photonic lattices~\cite{hartmann_strongly_2006,greentree_quantum_2006,angelakis_photon-blockade-induced_2007,schiro_phase_2012} and artificial gauge field systems~\cite{umucalilar_fractional_2012,keeling_superfluid_2011} to observe e.g. quantum-Hall-type physics. In contrast to present cold atom systems~\cite{bloch_many-body_2008}, quantum manybody ground states in such lattices are expected to be populated directly by thermalization. It will also be important to test for the phase coherence~\cite{snoke_dynamics_2013,leeuw_phase_2014} in detail and for superfluidity of the condensed photon gas~\cite{keeling_superfluid_2011}.

\begin{acknowledgments}
\vspace{2mm}
Financial support from the DFG (We1748-17) and the ERC (INPEC) is acknowledged.
\end{acknowledgments}

%\bibliographystyle{unsrt}   
\bibliography{dynbib.bib}
 
\end{document}